\documentstyle[12pt]{article}
\newcommand{\ald}{\dot \alpha}
\newcommand{\bed}{\dot \beta}

\newcommand{\dald}{{\bar{\partial}}_{\dot \alpha}}
\newcommand{\nald}{{\bar{\nabla}}_{\dot \alpha}}
\newcommand{\ned}{{\bar{\nabla}}_{\dot \beta}}

\newcommand{\ej}{\cal E}
\newcommand{\hej}{\hat {\ej}}
\newcommand{\beq}{\begin{equation}}
\newcommand{\eeq}{\end{equation}}
\begin{document}
\textwidth 160mm
\textheight 240mm
\topmargin -20mm
\oddsidemargin 0pt
\evensidemargin 0pt

\begin{titlepage}
\begin{center}

\huge{\bf Self-dual perturbiner in Yang-Mills theory}

\vspace{1.5cm}

\large{\bf K.G.Selivanov}

{ITEP, B.Cheremushkinskaya 25, Moscow, 117259, Russia}

\vspace{1.9cm}

{Contribution to Proceedings of the Europhysics Conference on HEP, Jerusalem'97}

\vspace{1.0cm}

\end{center}

 
 




\begin{abstract}
The {\it perturbiner} approach to the multi-gluonic amplitudes in Yang-Mills
theory is reviewed.
\end{abstract}
\end{titlepage}
\section{Definition and motivation}
{{\it Perturbiner} is a solution of field equations which can be defined in any field
theory \cite{RS1}.
{\it Definition.}
(I use for a 
moment the scalar field theory)
Consider linear part of the equations (that is, with no interactions)},
{and take its solution of the type of}
${{\phi}^{(1)}=\sum_{j}^{N}{\ej}_{j}}$,
{where ${\ej}_{j}=a_{j}e^{ik^{j}x}$, $k$'s are on-shell momenta, 
$k^{2}=m^{2}$ and $a_{j}$ is assumed to be nilpotent, $a_{j}^{2}=0$.
{The  perturbiner is a (complex) solution of the nonlinear field equations
which is: 
i)polynomial in ${\ej}_{j}$ with constant coefficients 
and
ii)whose first order in  ${\ej}_{j}$ part is precisely ${\phi}^{(1)}$}.

Generalization for higher spins is obvious: one should put a polarization factor in front
of ${\ej}_{j}$ in ${\phi}^{(1)}$.}\\
{\it Motivation.}
{The perturbiner so defined is a generating function for the tree form-factors.}
{The set of the plane waves entering the perturbiner is essentially the set of asymptotic 
states in the amplitudes
which the perturbiner is the generating function for. The nilpotency condition
 $a_{j}^{2}=0$ means that one considers only amplitudes with no multiple particles
in the same state.}\\
{\it Notice:}
i) due to this definition one actually works with a 
 finite-dimensional space of polynomials in $N$ nilpotent variables instead of
 an infinite-dimensional function space ;
ii)this definition is different from the one traditionally considered in the 
stationary phase approach to the $S$-matrix \cite{FS}.

{In general case, one cannot proceed in a different from the usual perturbation 
theory way. However in some theories and/or for some sets of the asymptotic states
included one can use other powerful methods.
In the Yang-Mills (YM) theory one can consider only the same helicity states
which leads to considering only the self-duality (SD) equations instead of the full
YM ones.
This type of SD solutions has been discussed
in refs.\cite{Ba}, \cite{Se}, \cite{KO}. In ref.\cite{Ba} and independently
in ref.\cite{Se}, the tree like-helicity amplitudes were related to solutions
of the SD equations. In \cite{Ba} it was basically shown that  the SD
equations reproduce the recursion relations for the tree form-factors
(also called one-gluonic currents) obtained originally in ref.\cite{BG}
from the Feynman diagrams; the  corresponding solution of SD equations
was obtained in terms of the solution of ref.\cite{BG}  
of the recursion relations for the ``currents''. In ref.\cite{Se} an example of SD 
perturbiner was
obtained in the SU(2) case by a 'tHooft anzatz upon further restriction on the
asymptotic states included. The consideration of ref.\cite{KO} is based
on solving  recursion relations analogous to refs. \cite{BG}.
In \cite{RS2} the YM SD perturbiner was constructed by the twistor methods
\cite{Ward}
which also allowed us to obtain perturbiner with one
opposite-helicity gluon and thus to obtain a generating
function for the so-called maximally helicity violating Parke-Taylor 
amplitudes \cite{PT},
\cite{BG}. In \cite{RS2} we also constructed the SD perturbiner in
the background of an arbitrary instanton solution.

Briefly, the twistor approach to the SD equations goes as follows. One introduces
an auxiliary twistor variable  $p^{\alpha}, \alpha=1,2$, which can be 
viewed on as a pair of complex numbers, and form objects
$A_{\ald}=p^{\alpha}A_{\alpha \ald},
\dald=p^{\alpha}\frac{\partial}{\partial x^{\alpha \ald}},
\nald=\dald+A_{\ald}$. In terms of $\nald$, the SD equations turn to a zero-curvature
condition, ${[\nald, \ned]=0, \; {\rm at \: any} \; p^{\alpha}, \alpha=1,2}$, which can be 
solved for $A_{\ald}=p^{\alpha}A_{\alpha \ald}$ as
\beq
\label{gauge}
{A_{\ald}=g^{-1}{\dald}g}
\eeq
{where $g$ is a function of $x^{\alpha \ald}$ and $p^{\alpha}$ with values in the 
complexification of the
gauge group. $g$ must depend on $p^{\alpha}$ in such a way that the resulting
$A_{\ald}$ is a linear function of $p^{\alpha}$, $A_{\ald}=p^{\alpha}A_{\alpha \ald}$ . 
Actually, $g$ is sought for as a homogeneous of 
degree zero rational function of $p^{\alpha}$. Such function necessary has singularities
in the $p^{\alpha}$-space and it is subject to condition of regularity of 
$A_{\ald}$. Then by construction, $A_{\ald}$ is
a homogeneous  of degree one regular rational function of two 
complex  variables $p^{\alpha}, \alpha=1,2$. As such, it is necessary
just linear in $p^{\alpha}$. 

An essential moment is that in the case of {perturbiner} $g^{ptb}$ can only be a
polynomial in the variables ${\ej}_{j}$. First order in ${\ej}_{j}$ term in $g^{ptb}$
is fixed by the plane wave solution of the free equation (that is by the set of asymptotic
states included), while the demand of regularity of $A_{\ald}$ fixes  $g^{ptb}$
up to the gauge freedom.

\section{The plane wave solution of the free equation}
{A solution of the free (i.e. linearized) SD equation consisting of $N$ plane waves 
looks as follows}
\beq
\label{free1}
{A^{(1) N}_{\alpha \ald}=
\sum_{j}^{N} \epsilon^{+j}_{\alpha \ald}{\hej}_{j}}
\eeq
{where the sum runs over gluons, $N$ is the number of gluons,\\
$\epsilon^{+j}_{\alpha \ald}$ is a four-vector defining a polarization of the $j$-th gluon,
${\hej}^{j}=t_{j}{\ej}^{j}=t_{j}a_{j}e^{ik^{j}x}$,
$t_{j}$ is a matrix defining color orientation of the $j$-th gluon.
$k^{j}_{\alpha \ald}$, as a light-like four-vector, decomposes into a product of two
spinors}
${k^{j}_{\alpha \ald}=\ae^{j}_{\alpha} \lambda^{j}_{\ald}}$.{\footnote{the reality 
of the four momentum in Minkowski space assumes that $\lambda_{\ald}=
{\bar {\ae}}_{\alpha}$}} 
{The polarization $\epsilon^{+j}_{\alpha \ald}$,
as a consequence of the linearized SD equations, also decomposes into a product
of spinors, such that the dotted spinor is the same as in the decomposition of momentum k,
$\epsilon^{+j}_{\alpha \ald}=\frac{q^{j}_{\alpha} \lambda^{j}_{\ald}}{( \ae^{j},q^{j})}$
where normalization factor  is defined with use of a convolution 
$(\ae^{j},q^{j} )=\varepsilon^{\gamma \delta}\ae^{j}_{\gamma}q^{j}_{\delta}=
{\ae^{j}}^{\delta}{q^{j}}_{\delta}$. Indexes are raised and lowered with the
$\varepsilon$-tensors.}
{The free anti-SD equation would give rise to a polarization
$\epsilon^{-}_{\alpha \ald}=\frac{\ae_{\alpha} \bar{q}_{\ald}}
{( \lambda,\bar{q})}$
The auxiliary spinors $q_{\alpha}$ and $\bar{q}_{\ald}$ form together a four-vector
$q_{\alpha \ald}=q_{\alpha}\bar{q}_{\ald}$ usually called a reference momentum.
The normalization  was chosen so that}
${\epsilon^{+} \cdot \epsilon^{-}=\varepsilon^{\alpha \beta}\varepsilon^{\ald \bed}
\epsilon^{+}_{\alpha \ald} \epsilon^{-}_{\beta \bed}=-1}$

\section{The solution for $g^{ptb}$ and  $A^{ptb}_{\ald}$}
{First order in ${\ej}$ terms in  $g^{ptb}$ are easily found from equation 
(\ref{gauge}), first order version of which reads}
${A^{ptb(1)}_{\ald}={\dald}g^{ptb(1)}}$ and hence, with use of 
(\ref{free1}), 
\beq
\label{regulator}
{g^{ptb(1)}=1+\sum_{j}\frac{(p,q^{j})}{(p, \ae^{j})}
 \frac{{\hat {\cal E}}^{j}}{(\ae^{j},q^{j})}}
\eeq
Thus $g^{ptb(1)}$ has simple poles on the auxiliary space at the points 
$p_{\alpha}={\ae}^{j}_{\alpha}$.
Actually, one can see that the condition of regularity of $A^{ptb}_{\ald}$
dictates that the full  $g^{ptb}$ has only simple poles at the same points as  $g^{ptb(1)}$.
Moreover, it also fixes residues of  $g^{ptb}$ to all orders in ${\ej}$ in terms of the 
residues of $g^{ptb(1)}$ Eq.(\ref{regulator}).

The known singularities of $g^{ptb}$ fix it up to an independent of the
auxiliary variables  $p^{\alpha}, \alpha=1,2$ matrix, i.e., up to a gauge freedom. 
The problem of reconstructing $g^{ptb}$ from its singularities essentially simplifies 
if one considers color ordered 
highest degree monomials in $g^{ptb}$,
for which one obtaines (\cite{RS2})
\beq
\label{bsol}
g^{ptb}_{N (N, \ldots ,1)}=\frac{(p, q^{N})(\ae^{N}, q^{N-1}) \ldots 
(\ae^{2}, q^{1})}{(p, \ae^{N})(\ae^{N}, \ae^{N-1}) \ldots
(\ae^{2}, \ae^{1})}
\frac{ {\hej}^{N}}{( \ae^{N},q^{N})} \ldots \frac{{\hej}^{1}}{( \ae^{1},q^{1})}
\eeq
This is, essentially, a solution of the problem.
Substituting $g^{ptb}$ (\ref{bsol}) into equation (\ref{gauge}) determines 
the perturbiner
$A^{ptb}_{\ald}$ (see ref.\cite{RS2}).


\section{The Parke-Taylor amplitudes}
The SD perturbiner can be used as a base point for a perturbation 
procedure of adding one-by-one gluons of the opposite helicity, or other particles,
say, fermions, interacting with gluons.. 
The explicit expression for $g^{ptb}$ (\ref{bsol}) is very useful 
in this procedure.
The SD perturbiner 
itself describes the tree form-factors  - objects including an arbitrary number of
on-shell SD gluons and one arbitrary off-shell gluon. To obtain the Parke-Taylor
amplitudes, those with
two gluons of the opposite helicity, one essentially needs to construct the perturbiner
including one on-shell gluon of the opposite helicity, 
that is, to solve the linearized YM equation 
in the background of SD perturbiner. Details of this solution can be found in 
ref.\cite{RS2}.
The resulting generating function for the Parke-Taylor amplitudes reads  
\begin{eqnarray}
M(k'',k', \{a_{j}\})=
{-i ( {\ae}'', \ae')^{2} \int d^{4}x \, tr} \nonumber\\ 
{{\hej}''(g^{ptb}|_{(p={\ae}', q={\ae}'')})^{-1} {\hej}'
g^{ptb}|_{(p={\ae}', q={\ae}'')}}
\end{eqnarray}
Considering cyclic ordered terms in this expression with $g^{ptb}$ (\ref{bsol}) one
easily reproduces the Parke-Taylor maximally helicity violating amplitudes
 \cite{PT}, \cite{BG}.

\section{The SD perturbiner in a topologically nontrivial sector}
The concept of perturbiner can be generalized
to a topologically nontrivial sector (see \cite{RS2}). In the latter case  it provides 
a framework for the 
instanton mediated  multi-particle amplitudes.
All what we need to know about the instanton, $A^{inst}_{\ald}$, that it can be represented
in the twistor-spirit form
$A^{inst}_{\ald}=g_{inst}^{-1} \dald g_{inst}$
$g_{inst}$ is assumed to be a rational function of  the auxiliary variables $p^{\alpha}$,
 such that $A^{inst}_{\ald}$ is a linear homogeneous function
of $p^{\alpha}$.  Then the SD topologically nontrivial perturbiner
$A^{iptb}_{\ald}$ is represented in the form 
$A^{iptb}_{\ald}=({g^{iptb}})^{-1} \dald g^{iptb}$
and the corresponding $g^{iptb}$ is found to be 
\beq
\label{iptb}
g^{iptb} ({\hej}^{1}, {\hej}^{2}, { \ldots }) =g^{inst}
g^{ptb} ({{\hej}^{1}}_{g}, {{\hej}^{2}}_{g}, { \ldots } )
\eeq
where ${\hej}^{j}_{g} $ stand for twisted harmonics
${\hej}^{j}_{g}=(g_{inst}|_{(p=\ae^{j})})^{-1} \hej^{j}g_{inst}|_{(p=\ae^{j})}$
and $g^{ptb}$ as in Eq.~(\ref{bsol}).
This is the sought for SD perturbiner in an arbitrary instanton background.

%


\begin{thebibliography}{99}
\bibitem{RS1}\ A.Rosly, K.Selivanov, preprint ITEP-TH-43-96,
 hep-th/9610070
\bibitem{FS}\ A.A.Slavnov, L.D.Faddeev, Introduction to the Theory of Quantum
 Qauge Fields,Nauka, Moscow, 1978
\bibitem{IZ}\ C.Itzykson, J.-B.Zuber, Quantum Field Theory, McGrow-hill, NY, 1980
\bibitem{Ba}\ W.Bardeen, Prog.Theor.Phys.Suppl, 123 (1996) 1
\bibitem{Se}\ K.Selivanov, preprint ITEP-21-96, hep-ph/9604206
\bibitem{KO}\ V.Korepin, T.Oota, J.Phys.A 29 (1996) 625, 
 hep-th/9608064
\bibitem{Ward}\ R.S.Ward, Phys.Lett.61A (1977) 81
\bibitem{PT}\ S.Parke, T.Taylor,  Phys.Rev.Lett. 56 (1986) 2459
\bibitem{BG}\ F.Berends, W.Giele, Nucl.Phys. B306 (1988) 759
\bibitem{RS2}\ A.Rosly, K.Selivanov, Phys.Lett.B 399 (1997) 135
\bibitem{mapa}\ M.Mangano, S.Parke, Phys.Rep. 200 (1991) 301

\end{thebibliography}
\end{document}